\documentclass[preprint2]{aastex}

\newcommand{\ba}{\begin{eqnarray}}
\newcommand{\ea}{\end{eqnarray}}
\newcommand{\be}{\begin{equation}}
\newcommand{\ee}{\end{equation}}
\newcommand{\ra}{\rangle}
\def\mpy{\rm \ M_\odot \ {\rm yr^{-1}}}
\def\rad{\rm \ rad \, {\rm m^{-2}}}

\slugcomment{to be submitted to \apj}

\shorttitle{Faraday Rotation from Simulations}
\shortauthors{Sharma \& Quataert}

\begin{document}

\title{Faraday Rotation in Global Accretion Disk Simulations: Implications for Sgr A*}

\author{Prateek Sharma, Eliot Quataert}
\affil{Astronomy Department, University of California,
    Berkeley, CA 94720}
\email{psharma@astro.berkeley.edu, eliot@astro.berkeley.edu}

\and

\author{James M. Stone}
\affil{Department of Astrophysical Sciences, Princeton University, Princeton, NJ 08544}
\email{jstone@astro.princeton.edu}

\begin{abstract}

We calculate Faraday rotation in global axisymmetric
magnetohydrodynamic simulations of geometrically thick accretion
flows.  These calculations are motivated by the measured rotation
measure (RM) of $\approx -6 \times 10^5$ rad m$^{-2}$ from Sgr A* in
the Galactic center, which appears to have been stable over the past
$\approx 7$ years.  In our numerical simulations, the quasi-steady
state structure of the accretion flow, and the RM it produces, depends
on the initial magnetic field threading the accreting material.  In
spite of this dependence, we can draw several robust conclusions about
Faraday rotation produced by geometrically thick accretion disks: i)
the time averaged RM does not depend that sensitively on the viewing
angle through the accretion flow, but the stability of the RM
can. Equatorial viewing angles show significant variability in RM
(including sign reversals), while polar viewing angles are relatively
stable if there is a large scale magnetic field threading the disk at
large radii. ii) Most of the RM is produced at small radii for polar
viewing angles while all radii contribute significantly near the
midplane of the disk.  Our simulations confirm previous analytic
arguments that the accretion rate onto Sgr A* must satisfy $\dot
M_{\rm in} \ll \dot M_{\rm Bondi} \sim 10^{-5} \mpy$ in order to not
over-produce the measured RM.  We argue that the steady RM $\approx -6
\times 10^5$ rad m$^{-2}$ from Sgr A* has two plausible explanations:
1) it is produced at $\sim 100$ Schwarzschild radii, requires
$\dot{M}_{\rm in} \approx 3 \times 10^{-8} M_\odot$ yr$^{-1}$, and we
view the flow at an angle of $\sim 30^\circ$ relative to the rotation
axis of the disk; in our simulations, the variation in RM across a
finite-sized source is sufficient to depolarize the emission below
$\approx$ 100 GHz, consistent with observations. 2) Alternatively, the
RM may be produced in the relatively spherical inflowing plasma near
the circularization radius at $\sim 10^3-10^4$ Schwarzschild radii,
with the magnetic field perhaps amplified by the magnetothermal
instability. Time variability studies of the RM can distinguish
between these two possibilities.

\end{abstract}

\keywords{accretion, accretion disks -- MHD -- Galaxy: center}

\section{Introduction}
Observations of stellar orbits in the Galactic center indicate the
presence of a $\approx 4 \times 10^{6} M_\odot$ black hole at the
center of the Galaxy \citep[e.g.,][]{sch02,ghe05}. Numerical modeling
of stellar outflows and X-ray observations of the diffuse thermal
plasma in the central parsec show that winds from massive stars
provide a mass supply of $\dot M_{\rm Bondi} \sim 10^{-6}-10^{-5}
M_\odot$ yr$^{-1}$ at $\approx 10^5$ Schwarzschild radii, which is
approximately the gravitational sphere of influence of the black hole
\citep{bag03,qua04,cua06}.  The observed bolometric luminosity of the
electromagnetic source Sgr A* coincident with the black hole is,
however, 5 orders of magnitude smaller than would be expected if all
of this available gas were accreted with a radiative efficiency of
10\%.  As a result, the inflowing plasma must have a very low
radiative efficiency (e.g., \citealt{nar95}) and/or very little of the
gas supplied at large radii actually makes it down to the black hole
(e.g., \citealt{bla99}).  Global MHD simulations of radiatively
inefficient accretion flows show significant mass-loss consistent with
the latter possibility (\citealt{sto01}, hereafter SP01;
\citealt{haw01}).  In addition, local numerical simulations of
electron heating in collisionless accretion flows find that the
electrons receive a significant fraction of the gravitational
potential energy released in the accretion flow, requiring $\dot{M}
\ll \dot{M}_{\rm Bondi}$ to explain the low luminosity of Sgr A*
\citep[][]{sha07}.

An important observational diagnostic of the physical conditions in
the accreting plasma is polarization measurements.  Linear
polarization from Sgr A* was first detected by~\citet{ait00} at
sub-millimeter wavelengths; they found that the polarization fraction
increases rapidly from 150 to 400 GHz.  Linear polarization is not
detected at lower frequencies, although circular polarization is
(e.g., \citealt{bow99,bow02}). The decrease in the linear polarization
fraction at low frequencies may be due to depolarization by
differential Faraday rotation across the source (e.g.,
\citealt{bow99}).  In addition, the emission at low frequencies is
optically thick, which suppresses the linear polarization if the
particle distribution function is roughly thermal (e.g.,
\citealt{goldston}).

The detection of linearly polarized radiation at mm \& sub-mm
wavelengths rules out high accretion rate models in which $\dot M \sim
\dot M_{\rm Bondi}$ near the black hole \citep{ago00,qua00,melia}.
Analytic estimates show that such models would produce a rotation
measure (RM) much larger than observed---so large, in fact, that all
of the mm emission would be depolarized by Faraday rotation.
In the past five years, higher resolution constraints on the linear
polarization of Sgr A* have generally confirmed the Aitken et
al. results \citep[][]{bow03,mar06,mac06}.  \citet{mar07} performed
the first measurement of the polarization of Sgr A* with sufficient
sensitivity to determine the Faraday rotation in single-epoch
observations, free from ambiguities introduced by possible time
variation in the intrinsic polarization angle.  They found $RM
\approx -6 \times 10^{5}$ rad m$^{-2}$ using observations at both 227
and 343 GHz.  The observed RM appears to be relatively constant over
the 2 years of high sensitivity SMA observations (Marrone, private
communication).  Comparison with previous observations suggests that
the RM towards Sgr A* has not changed sign over the past $\approx 7$
years.

Propagation through the interstellar medium cannot account for the
large RM observed towards Sgr A*.
In addition, X-ray observations of the thermal plasma in the central
parsec of the Galactic center shows that the electron number density
is $\approx 100$ cm$^{-3}$ and the temperature is $\approx 2$ keV at a
distance of $\approx 1" \approx 0.04 {\rm pc} \, \approx 10^5 r_g$
from the black hole \citep{bag03,qua04}, where $r_g$ is the
Schwarzschild radius.  If the magnetic field on this scale were
uniform, purely radial, and in equipartition with the thermal
pressure, $B \approx 2$ mG and $RM \approx 5000$ rad m$^{-2}$, two
orders of magnitude smaller than what is observed.  This shows that
the observed RM must be produced at distances $\ll 10^5 r_g$ from the
black hole, providing an important diagnostic of the physics of
accretion onto Sgr A*.  In view of the uniqueness of this probe, and
the importance of Sgr A* as the paradigm for radiatively inefficient
accretion flows, a more detailed study of the constraints imposed by
the observed Faraday rotation is warranted.

In this paper we carry out two-dimensional (axisymmetric) MHD
simulations of accretion onto a central point mass and study Faraday
rotation through the resulting turbulent accretion flow.  This
significantly improves on previous analytic estimates of the Faraday
rotation in radiatively inefficient accretion flows, which relied on
simplifying assumptions such as power-law density profiles, uniform
steady magnetic fields, etc. One caveat about our assumption of
axisymmetry is that, by the anti-dynamo theorem, the magnetic field
must eventually decay. However, the similarity of previous 2-D and 3-D
simulations in the quasi-steady turbulent state, that lasts for
hundreds of orbital periods at small radii \citep[][]{haw01}, gives us
some confidence that our results will hold in 3-D.

The remainder of this paper is organized as follows.  Our numerical
techniques and methods for calculating RM are described in \S 2; we
also briefly summarize the analytic scalings for RM in radiatively
inefficient accretion flows (\S \ref{sec:analytic}).  In \S 3 we
discuss our results on Faraday rotation in MHD simulations.  We focus
on a fiducial model, but discuss the sensitivity of our results to the
initial magnetic field configuration in the simulation, numerical
resolution, and the size of the computational domain. In \S 4 we
summarize our results and discuss several scenarios that can account
for the observed RM from Sgr A*.

\section{Methods and Analytic Scalings}

\subsection{Numerical Simulations}

Unless specified otherwise, we use a numerical set up similar to that
of SP01. We use the ZEUS-2D MHD code \citep[][]{sto92a,sto92b} with
spherical ($r,\theta$) coordinates to solve the equations of ideal MHD
under the assumption of axisymmetry.  The gravitational potential is
the pseudo-Newtonian potential of~\citet{pac80}, namely $\Phi = -
GM/(r-r_g)$, where $r_g=2GM/c^2$.  The energy equation is adiabatic
with the exception of heating by shock-capturing artificial viscosity
(which is a rather minor source of heating).  We do not use an
explicit resistivity to capture some of the energy lost to grid-scale
averaging as it does not appear to significantly affect the dynamics
of the flow (e.g., see SP01).

The initial condition is a geometrically thick, weakly magnetized,
rotationally and pressure supported constant angular momentum torus.
The density in the initial torus is given by equation (6) of
\citet{sto99} which is based on \citet{pap84}.  The peak density of
the torus is $1$ in code units.  The torus is truncated when the
density falls below 0.01.  It is surrounded by a constant density
corona ($\rho=10^{-4}$); there is a density and temperature jump at
the torus-corona boundary.

The torus is specified by the radius of the initial density maximum,
$r_0$, which we take to be $r_0=40r_g$, $100r_g$, or $200r_g$.  The
gravitational sphere of influence of Sgr A* on the surrounding hot
plasma is $\sim 10^5 r_g$.  Three-dimensional numerical simulations of
accretion from stellar winds imply a circularization radius of $\sim
10^3-10^4 r_g$ \citep[][]{cua06}. Ideally we would thus like to
simulate an accretion disk which is $\sim 10^4-10^5 r_g$ in radial
extent. This is, however, infeasible at the present time.  Instead,
our goal is to understand the generic properties of Faraday rotation
through the steady state accretion flow structure that develops at
radii $\lesssim r_0$.

Our grid is logarithmic with a higher density of points at small radii
and in the midplane.  The radial grid is spaced such that the number
of grid points from $r_{\rm min}$ to $(r_{\rm min}r_{\rm max})^{1/2}$
is roughly the same as the number of grid points from $(r_{\rm
min}r_{\rm max})^{1/2}$ to $r_{\rm max}$, where $r_{\rm min}$ and
$r_{\rm max}$ are the inner and outer radial boundaries.  Three
different resolutions, $60\times 44$, $120 \times 88$, and $240 \times
176$, are used.  We use outflow boundary conditions and set $B_r B_\phi
\leq 0$ at $r_{\rm min}=2 r_g$.

We consider two different initial magnetic field configurations in the
torus: configuration A in which the field lies along the constant
density contours (Fig. \ref{fig:fig1}) and is derived from the vector
potential $A_\phi=\rho^2/\beta_0$ with $\beta_0=200$ (the same as runs
B \& F of SP01), and configuration B which has a net radial field in
the equatorial plane (see Fig. \ref{fig:fig8} \& the Appendix). We find
that the results of our Faraday rotation calculations remain sensitive
to the initial field configuration even in the turbulent state.  This
is discussed in more detail in \S 3.2 and \S 4.

The initial weakly magnetized torus is unstable to the
magnetorotational instability (MRI; \citealt{bal91}), which rapidly
amplifies the magnetic field leading to MHD turbulence and efficient
angular momentum transport. After an initial amplification stage, a
quasi-steady turbulent accretion flow is established. However, at late
times, due to the impossibility of a self sustained dynamo in 2D, the
turbulent motions die away. We carry out the RM calculations in the
quasi-steady turbulent state which lasts for hundreds of orbital
periods at small radii.

\subsection{Rotation Measure}

\label{sec:RM}

Due to the different index of refraction for left and right circularly
polarized radiation, the polarization angle of linearly polarized
light is rotated by an angle of $\lambda^2$RM on passing through a
non-relativistic magnetized plasma, where RM $= (e^3/2\pi m_e^2c^4)
\int n {\bf B \cdot dl}$ is the rotation measure, $e$ and $m_e$ are
the electron charge and mass, $c$ is the velocity of light in vacuum,
$n$ is the number density of electrons, ${\bf B}$ is the magnetic
field, and ${\bf dl}$ is the differential length along the line of
sight.  For a relativistic plasma, RM is suppressed by a factor of
$\approx \ln(\theta)/\theta^2$ (\citealt{qua00}), where $\theta =
kT_e/m_e c^2$.  In addition, propagation through a relativistic plasma
can lead to conversion between linear and circular polarization (e.g.,
Ruszkowski \& Begelman 2002; Ballantyne, Ozel, \& Psaltis 2007).  We
focus on non-relativistic plasmas throughout this paper.

We calculate RM along different lines of sight through the simulation
domain in the quasi-steady turbulent state.  The MHD variables
($\rho$, $B$, etc.) are output 40 times per orbital period at the
density maximum $r_0$; these variables are then used to calculate RM
for different viewing angles.  For Figure \ref{fig:fig7} we used
output 160 times per orbital period to guarantee that the short
timescale variability in RM was captured.  Our results are, however,
largely insensitive to the exact time sampling used.

The rotation measure in arbitrary code units is rescaled as follows
\be
\label{eq:RMcode} RM_{\rm code} \equiv
\left ( \frac{\langle \dot{M}_{\rm code}\ra}{0.002} \right )^{-3/2}
\times \int_{r_{\rm in}}^{r_{\rm out}} \rho {\bf B \cdot dl} \ee where
$\langle \dot{M}_{\rm code} \ra$ is the average mass accretion rate at
$r_{\rm min} = 2 r_g$ during the turbulent steady state, and $r_{\rm
in}$ and $r_{\rm out}$ are the inner and outer radii used in
calculating the rotation measure.  Numerical simulations with
different resolution, $r_0$, and initial field configurations have
different values of $\langle \dot M_{\rm code} \ra$.  Equation
(\ref{eq:RMcode}) effectively re-scales all of the different
simulations to have the same time averaged accretion rate, allowing a
more useful comparison of the different simulations.  The conversion
of $RM_{\rm code}$ to a physical rotation measure (in rad m$^{-2}$) is
then given by 
\be
\label{eq:RM}
\nonumber RM = 1.5 \times 10^{10} \left ( \frac{\dot{M}_{\rm
in}}{10^{-5}M_\odot {\rm yr}^{-1} } \right )^{3/2} \left(
\frac{M_{\rm BH}}{4 \times 10^6 M_\odot} \right )^{-2} \left (
\frac{RM_{\rm code}}{10^{-4}} \right ),
\ee
where $M_{\rm BH}$ is the
black hole mass and $\dot{M}_{\rm in}$ is the actual mass accretion
rate through the horizon of the black hole (i.e., at $r_{\rm min} = 2
r_g$).

Physically $r_{\rm in}$ in equation (\ref{eq:RMcode}) corresponds to
the radius where the electrons become non-relativistic, since there is
little Faraday rotation from smaller radii.  We choose $r_{\rm in}=10
r_g$ unless specified otherwise.  The true $r_{\rm in}$ in Sgr A* is
probably somewhat larger, $r_{\rm in} \sim 100 r_g$ (e.g., Sharma et
al. 2007), but we are restricted to smaller values of $r_{\rm in}$ for
numerical reasons.  We discuss the scalings of our results to larger
$r_{\rm in}$ in \S 4.  The outer radius $r_{\rm out}$ is chosen so
that the density structure of the accretion flow is not affected by
the initial conditions. We choose $r_{\rm out}=30 r_g$ for
$r_0=40r_g$, $r_{\rm out} = 60 r_g$ for $r_0=100r_g$, and $r_{\rm out}
= 120 r_g$ for $r_0=200r_g$.

We calculate the RM in two ways in the simulations: along individual
lines of sight ($\theta=$ constant in spherical coordinates) and in a
cylindrical ``beam'' of diameter $D_{\rm beam}$.  Physically, the
relevant beam size is expected to be the size of the synchrotron
source which is itself a function of frequency, with $D_{\rm beam}
\approx 10 r_g (\nu/100 \, {\rm GHz})^{-1}$ for Sgr A*
\citep{bow04,shen05,lw07}.  We take $D_{\rm beam}=10r_g$ but find that
our results are not sensitive to a factor of $\sim 2$ variation in
this choice.

For an individual line of sight we evaluate the integral in equation
(\ref{eq:RMcode}) by adding up the contribution of each $\theta$ =
constant grid point that lies between $r_{\rm in}$ and $r_{\rm
out}$. For a beam of width $D_{\rm beam}$ all grid points lying within
the beam contribute to the RM integral. This effectively assumes that
the source has a constant surface brightness across the synchrotron
photosphere. Since the size of the grid cells is not uniform,
we calculate the beam averaged RM by calculating the average value of
$n B_\parallel$ within the beam, weighing the contribution of each
grid cell by its area, and multiplying by $r_{\rm out}-r_{\rm in}$
(the radial length of the region contributing to the calculated RM).
Using the RM for each time-slice in the turbulent state we
calculate the time averaged RM and the standard deviation in time, for
both individual lines of sight and beams.  The standard deviation in
time provides a useful measure of the time variability of the RM.  We
also calculate the spatial variation in RM within the beam of width
$D_{\rm beam}$; this can be used to quantify the effects of beam
depolarization.

\subsection{Analytic Scalings}

\label{sec:analytic}

To provide some context for the numerical simulations that follow, we
briefly review the analytic expectations for RM in radiatively
inefficient accretion flows.  For a flow with a density profile of $n
\propto r^{-3/2 + p}$ and a virial temperature $T \propto r^{-1}$, the
equipartition magnetic field strength should vary as $B \propto (n
T)^{1/2} \propto r^{-5/4 + p/2}$.  Thus the local contribution of a
given decade in radius to the rotation measure is given by $d RM/d \ln
r \propto r n B \propto r^{-7/4 + 3p/2}$.  For Bondi accretion, $p =
0$ and $d RM/d \ln r \propto r^{-7/4}$; RM is thus dominated by small
radii.  For Sgr A*, the Bondi scalings may be applicable outside the
circularization radius of $\sim 10^3-10^4 r_g$ \citep[][]{cua06},
although this need not be true in the presence of magnetic fields or
significant heat conduction (e.g., \citealt{in,jq}).

In contrast to the Bondi model, analytic models \citep{bla99,qua00a}
and numerical simulations \citep{sto01,haw01} of rotating radiatively
inefficient accretion flows find $p \sim 1$, which corresponds to an
accretion rate which decreases with radius as $\dot M \propto r$.  In
this case, $d RM/d \ln r \propto r^{-1/4}$.  There is thus a
significant contribution to RM from both small and large radii in the
flow.

\section{Numerical Results}

\subsection{The Fiducial Simulation}

\label{sec:fid}

\begin{figure}
\centering
\plotone{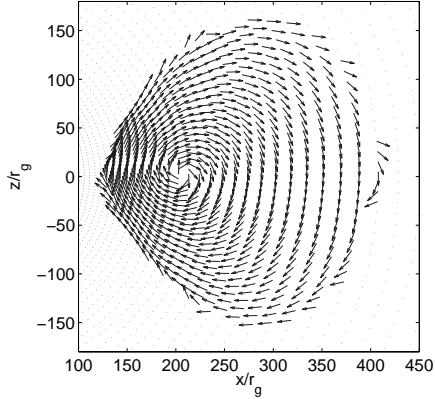}
\caption{Magnetic field unit vectors for initial field configuration A.
The initial magnetic field is confined within the dense torus and is
along constant density contours.
\label{fig:fig1}}
\end{figure}

As a fiducial case, we focus on the simulation with $r_0 = 200r_g$, a
resolution of $120 \times 88$ grid points, and with initial magnetic
field unit vectors shown in Figure \ref{fig:fig1} (``configuration
A'').  The resolution in the innermost portions of the torus is such
that the fastest growing MRI mode is resolved with about 5 grid
points.  Different resolutions, $r_0$, and initial magnetic field
configurations are discussed later.  Several properties for these
simulations are summarized in Tables \ref{tab:tab1} \& \ref{tab:tab2}.

\begin{figure}
\centering
\plotone{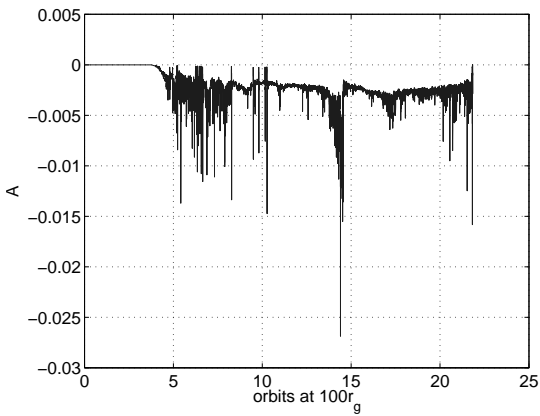}
\caption{The mass accretion rate through the inner boundary ($r_{\rm
min}=2r_g$) for the fiducial run. There is large variability in the
accretion rate on timescales comparable to the orbital period at small
radii. \label{fig:fig2}}
\end{figure}

The shear in the initial torus converts radial magnetic field into
azimuthal field. Simultaneously, the MRI amplifies the magnetic
field. The dynamical timescales are shortest at small radii in the
torus, where mass flows in as it loses angular momentum because of
magnetic stresses.  Figure \ref{fig:fig2} shows the mass accretion
rate through $r_{\rm min} = 2 r_g$ as a function of time for this
simulation.  A quasi-steady turbulent accretion flow is established
after 5 orbital periods at 100 $r_g$ (note that 1 orbital period at
100 $r_g$ is $\approx 3 \, (M_{\rm BH}/4 \times 10^6 M_\odot)$ days).
The accretion rate is, however, highly variable and can sometimes
change by an order of magnitude on short time scales (of order the
orbital timescales at small radii).  This is somewhat reminiscent of
the flaring activity seen in Sgr A* \citep[][]{bag01,gen03}.

\begin{figure}
\epsscale{1.11}
\centering \plottwo{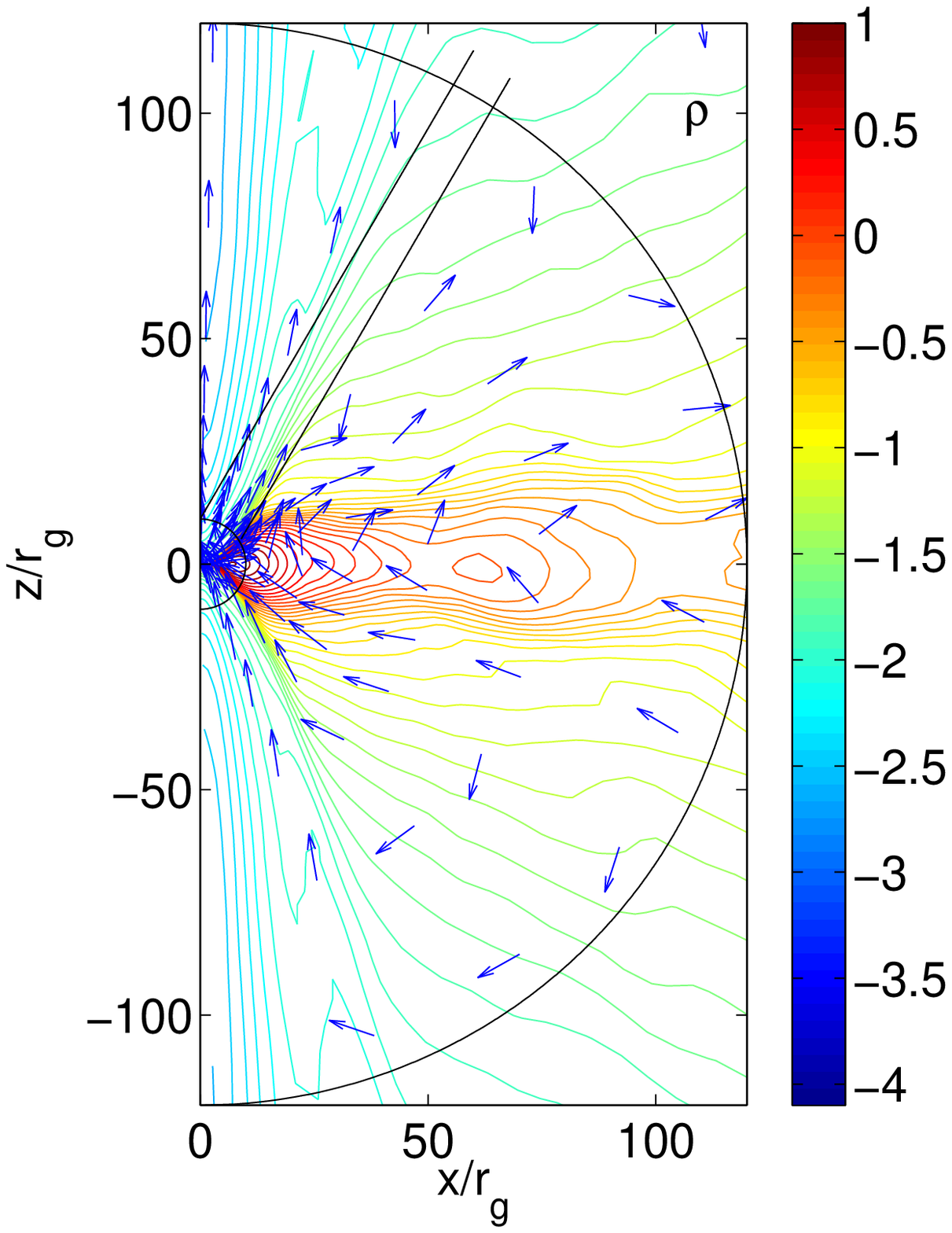}{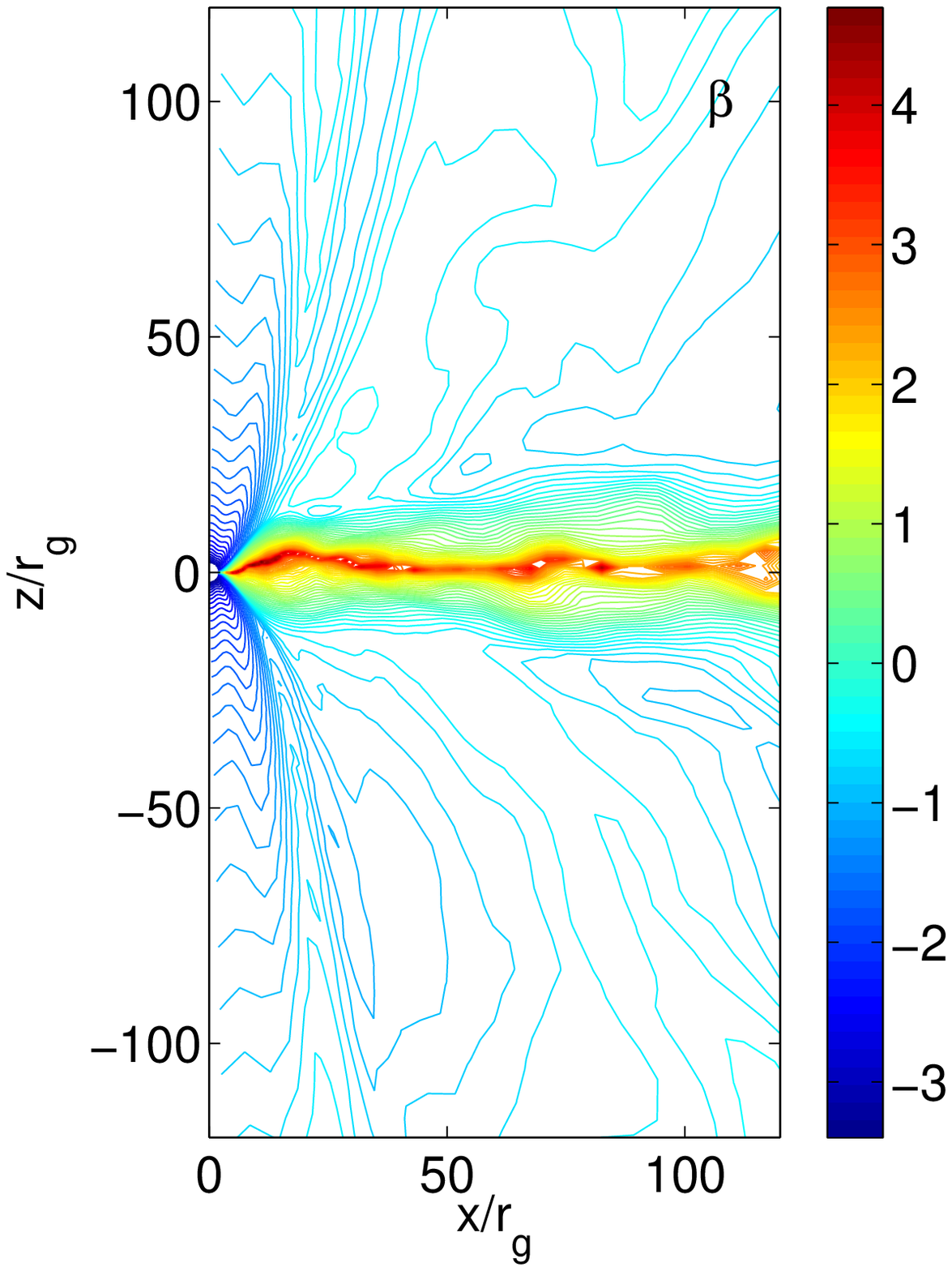}
\caption{Contour plots of density ($\rho$, left) and $\beta=8\pi
p/B^2$ (right) in the quasi-steady turbulent state for the fiducial
run (averaged from 6 to 20 orbits at $100r_g$).  Arrows indicate the
average magnetic field direction.  Dense high $\beta$ plasma is
confined to a relatively thin disk; the density decreases
significantly towards the poles, where the plasma is strongly
magnetized (low $\beta$). The inner ($r_{\rm in}=10 r_g$) and outer
($r_{\rm out}=120 r_g$) radii used in our rotation measure integrals
are shown.  Also shown is a fiducial ``beam'' with a diameter of
$D_{\rm beam}=10r_g$ at $\theta=30^\circ$ used to calculate the rotation
measure for a finite sized source.\label{fig:fig3}}
\end{figure}

Figure \ref{fig:fig3} shows contour plots of the density and plasma
$\beta$ (the ratio of thermal to magnetic pressure) for the fiducial
simulation, averaged over the quasi-steady turbulent state.  Poloidal
magnetic field unit vectors are superimposed on the density contours.
Figure \ref{fig:fig3} shows that a high density, moderately thin disk
with very high $\beta$ is formed in the equatorial plane of the initial
torus (similar to Figures 4 \& 5 of SP01). The accretion flow in the
turbulent state is influenced by the initial magnetic field
configuration.  Shearing of the initially radial magnetic field above
and below the equator leads to a strong toroidal field in the corona
which compresses the midplane into a dense moderately thin disk with
high $\beta$.
By contrast, both the density and $\beta$ are quite small in the polar
regions.  The average magnetic field vectors shown in Figure
\ref{fig:fig3} for the quasi-steady structure clearly maintain memory
of the initial magnetic field configuration shown in Figure
\ref{fig:fig1}.

\begin{figure}
\plotone{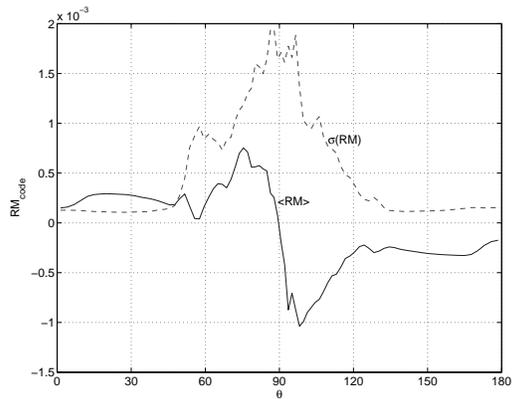}
\caption{Time averaged RM (solid line) and standard deviation in time
(dashed line) along individual lines of sight ($\theta=$ constant) in
the turbulent steady state for the fiducial run ($r_{\rm in}=10r_g$,
$r_{\rm out}=120r_g$).  Temporal fluctuations in RM in the turbulent
disk ($\sim 10^\circ$ about the midplane) are much larger than the
average RM. Only in the polar regions ($\sim 40^\circ$ about the
poles) are the fluctuations in RM smaller than the
mean.\label{fig:fig4}}
\end{figure}

For our fiducial simulation, we use $r_{\rm in}=10 r_g$ and $r_{\rm
out}=120 r_g$ as the inner and outer radii in the RM integral (eq.
[\ref{eq:RMcode}]). Figure \ref{fig:fig4} shows the time averaged
$RM_{\rm code}$ and the standard deviation in time along different
lines of sight.  As expected, the RM is fairly symmetric about the
midplane ($\theta=90^\circ$), except for the sign change which arises
because of the initial field configuration. In contrast to $\langle
RM_{\rm code} \ra$, the standard deviation peaks at the midplane and
is significantly smaller near the pole.  This is because the midplane
is highly turbulent, while the polar regions contain a comparatively
stable magnetic field.
Figure \ref{fig:fig4} shows that the time variability in RM is larger
than the average RM except within $\approx 40^\circ$ of the poles.

\begin{figure}
\plotone{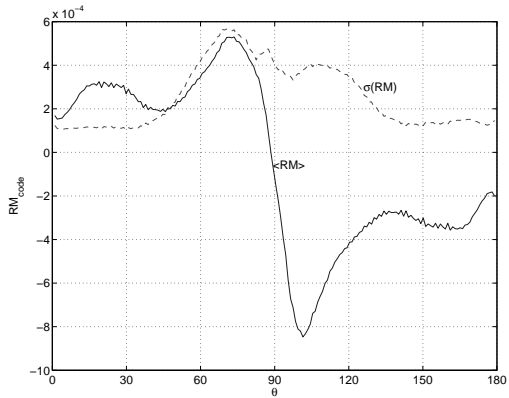}
\caption{Time averaged RM (solid line) and standard deviation in time
(dashed line) in a $10 r_g$ diameter beam as a function of viewing
angle for the fiducial run ($r_{\rm in}=10r_g$, $r_{\rm out}=120r_g$).
Fluctuations near the midplane of the disk ($\theta \sim 90^\circ$)
are somewhat smaller than for the individual lines of sight shown in
Figure \ref{fig:fig4}.  Nonetheless, only near the poles ($\theta \sim
30^\circ$) are the fluctuations smaller than the
mean.\label{fig:fig5}}
\end{figure}

As discussed in \S \ref{sec:RM}, a more apt comparison to observations
is the Faraday rotation averaged over a finite source size, rather
than along individual lines of sight.  Figure \ref{fig:fig5} shows
$RM_{\rm code}$ and the standard deviation as a function of $\theta$
for $D_{\rm beam} = 10 r_g$.  The results in Figure \ref{fig:fig5} are
reasonably similar to the single angle results shown in Figure
\ref{fig:fig4}.  The primary difference is that the fluctuations in
the RM in the midplane ($\theta \sim 90^\circ$) are smaller by a
factor of $\sim 3$ because the beam averages over the spatial
fluctuations in the turbulent disk.  Nonetheless, the temporal
fluctuations implied by Figure \ref{fig:fig5} are still sufficiently
large that one would only expect a relatively steady RM for polar
viewing angles.  Note also that both $\langle RM_{\rm code} \rangle$
and its standard deviation do not depend that strongly on viewing
angle $\theta$; there is roughly a factor of $\sim 5$ variation from
pole to equator.

\begin{figure}
\plotone{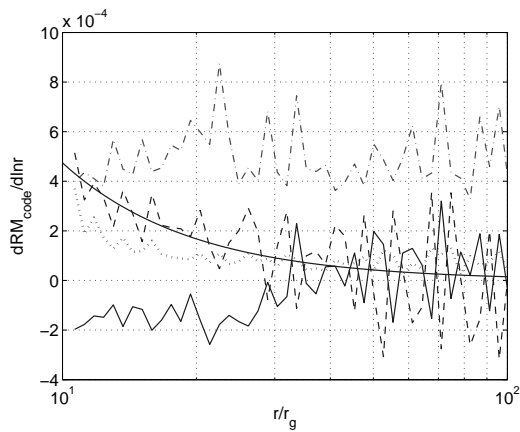}
\caption{$dRM_{\rm code}/d\ln r$ for a $10 r_g$ beam for the fiducial
simulation. Time average (solid line) and standard deviation in time
(dot-dashed line) for an equatorial viewing angle and time average
(dashed line) and standard deviation in time (dotted line) for a
$30^\circ$ viewing angle are shown. Also shown is an $r^{-3/2}$ fit to
$\langle dRM_{\rm code}/d\ln r \ra$ for $\theta=30^\circ$.  The curves
are oscillatory in part because $dRM_{\rm code}/d\ln r$ requires
taking a numerical derivative.
\label{fig:fig6}}
\end{figure}

To quantify which radii dominate the contribution to the RM, it is
useful to consider the logarithmic derivative $dRM/d \ln r$. Figure
\ref{fig:fig6} shows $dRM_{\rm code}/d\ln r$ as a function of radius
for an equatorial and $\theta=30^\circ$ beam with $D_{\rm beam}=10
r_g$; both the time average and standard deviation are
shown. $dRM_{\rm code}/d\ln r$ fluctuates with radius in Figure
\ref{fig:fig6} because we are taking a numerical derivative with a
finite number of grid points in the beam.  For the equatorial line of
sight, both the time average (solid) and standard deviation
(dot-dashed) of RM receive roughly equal contributions from all radii.
This is consistent with the analytic scalings reviewed in \S
\ref{sec:analytic}.  In addition, as already shown in Figure
\ref{fig:fig5}, the standard deviation is significantly larger than
the time averaged RM for the equatorial beam, i.e., the RM is highly
variable in the turbulent disk.  By contrast, the $\theta=30^\circ$
beam has a $\langle dRM_{\rm code}/d\ln r \rangle$ (dashed) which is
somewhat larger than the standard deviation (dotted).

For the polar line of sight through the corona of the disk, both
$\langle dRM_{\rm code}/d\ln r \ra$ and the standard deviation are
dominated by small radii, in contrast to the equatorial lines of sight
where all radii contribute significantly.  The solid line superimposed
on the numerical results in Figure \ref{fig:fig6} is a $dRM_{\rm
code}/d\ln r \sim r^{-3/2}$ power law which does a reasonable job at
fitting the numerical results.

Spatial variations in the amount of Faraday rotation across a
finite-sized source can lead to depolarization of the source at
wavelengths where $\sigma_\perp(RM) \lambda^2 \gtrsim 1$, where
$\sigma_\perp(RM)$ is the spatial standard deviation in RM across the
source (``beam depolarization''; e.g., \citealt{bur66,tri91}).  To
quantify this effect in our simulations, Figure \ref{fig:fig7} shows
the beam averaged RM and $\sigma_\perp(RM)$ as a function of time for
$\theta=30^\circ$ and $\theta=90^\circ$, taking $D_{\rm beam}=10r_g$
and $r_{\rm in}=10r_g$; $\sigma_\perp(RM)$ is calculated by
subdividing the $10 r_g$ wide beam into smaller beams.  The time
average value of $\sigma_\perp(RM)/RM$ is $\approx 10$ for
$\theta=90^\circ$ and $\approx 0.3$ for $\theta=30^\circ$; it is
somewhat smaller for $\theta < 30^\circ$.  For a larger value of
$r_{\rm in} = 25r_g$ we find similar results for $\theta=90^\circ$ but
find that the spatial variation in RM is somewhat larger at
$\theta=30^\circ$, with an average value of $\sigma_\perp(RM)/RM
\approx 0.9$.

\subsection{Effects of the Initial Magnetic Field Configuration}

The results of the previous section show that several properties of
the quasi-steady turbulent state in the accretion flow at late times
are determined by the initial magnetic field threading the torus at
the beginning of the simulation.  In particular, the antisymmetry of
RM about the equatorial plane (Fig. \ref{fig:fig5}) and the relatively
coherent magnetic field at polar angles (Fig. \ref{fig:fig3} and
\ref{fig:fig5}) are all signatures of the initial magnetic field
structure shown in Figure \ref{fig:fig1}.  It is thus important to
understand which of our results on the disk structure and the RM are
robust to changes in the initial field configuration.  Figure
\ref{fig:fig8} shows the initial magnetic field unit vectors for the
alternative initial field configuration we considered (``configuration
B''; see the Appendix for the vector potential used to generate this
field). In this case, the initial field is symmetric about the
midplane and has a net radial magnetic field in the equatorial plane
of the initial torus; $\beta$ is initially $\approx 10-30$ in the
equatorial plane and $\approx 100-10^3$ in the rest of the torus.
Note that although the initial magnetic field is still confined to the
torus, the field lines are not along the constant density contours (as
they were in Fig. \ref{fig:fig1}). In addition, the vertical scale of
the magnetic field structure in configuration B is roughly half that
of configuration A. Thus we need a higher resolution to resolve the
dynamics of the MRI.  We take the simulation with $r_0=100r_g$ and a
resolution of $240 \times 176$ as the standard case for initial field
configuration B. For this case the fastest growing MRI modes at $120
r_g$ (just above and below the equator) are resolved with about 8 grid
points.

Figure \ref{fig:fig9} shows 2-D contour plots of the density and
plasma $\beta$ averaged over the turbulent steady state. The resulting
disk is significantly thicker compared to that arising from field
configuration A (see Fig. \ref{fig:fig3}).  In addition, the disk
scale-height is roughly proportional to the radius as in analytic
models of radiatively inefficient accretion flows. Unlike
configuration A where $\beta \gg 1$ in the disk midplane, here $\beta
\sim 1$ in the turbulent disk.  The average field structure in the
quasi-steady state is again reminiscent of the initial field
configuration with radially inward field lines in the equatorial
plane.  The thick disk is surrounded by a relatively high $\beta$
turbulent region, with fairly low $\beta$ plasma at the poles.  In
general, the density and magnetic field structure of the disk is
rather different from that in Figure \ref{fig:fig3}.  This is also
shown in Figures \ref{fig:fig10} \& \ref{fig:fig11}, which compare
the density and $\beta$ within $12^\circ$ of the equatorial plane
averaged over the turbulent state as a function of radius for the two
different initial magnetic configurations.  The magnetic field is
stronger for configuration B, and $\beta$ decreases with radius,
likely because of flux freezing of the initial radial field.  In
addition, the density does not increase significantly inwards for
configuration B.  These differences in the steady state disk structure
appear be a direct consequence of the different initial magnetic field
structure.  Although a net inward radial field may not always be a
realistic initial condition, part of the motivation for choosing this
particular configuration was to understand whether the RM in the bulk
of the disk showed significant time variability even in the extreme case
of a coherent initial radial field.

Figure \ref{fig:fig12} shows the time averaged RM and the standard
deviation with $D_{\rm beam}=10 r_g$ as a function of angle $\theta$.
The RM is roughly symmetric about the equatorial plane and peaks at
$\theta \approx 90^\circ$. This is a direct consequence of the initial
magnetic field configuration, which is symmetric about the equator and
has a net radial field in the equator of the torus.  Figure
\ref{fig:fig12} also shows that the time variations in RM are larger
than $\langle RM_{\rm code} \ra$ at essentially {all angles}.  This is
in contrast to the results for initial magnetic field configuration A
shown in Figure \ref{fig:fig5} where $\langle RM_{\rm code} \ra$ is
larger than the fluctuations near pole. In the present case, the polar
viewing angles are relatively turbulent and show magnetic field
reversals in time.  Interestingly, although $\langle RM_{\rm
code} \ra$ is non-zero in the equator, the fluctuations due to the
turbulence are still sufficiently large that one would expect
significant time variability and sign changes in RM for equatorial
viewing angles.

The time averaged $dRM_{\rm code}/d \ln r$ and its standard deviation
for equatorial and $\theta=30^\circ$ beams with $D_{\rm beam}=10 r_g$
are shown as a function of radius in Figure \ref{fig:fig13}.  For the
polar viewing angle, the smallest radii dominate the contribution to
both $\langle RM_{\rm code} \ra$ and its time variation, with
$dRM_{\rm code}/d \ln r \propto r^{-1}$ being a reasonable fit.  For
the equatorial beam, however, both $\langle RM_{\rm code} \ra$ and the
standard deviation are dominated by the largest radii.  This is in
contrast to the results shown in Figure \ref{fig:fig6} for initial
field configuration A, and in contrast to the simple analytic
estimates from \S \ref{sec:analytic}. The significant contribution of
large radii to $\langle RM_{\rm code} \ra$ in the equatorial plane is
in part because the density in the equator increases with radius
for initial field configuration B (see Fig. \ref{fig:fig10}).

\subsection{Convergence Studies}

To test whether our results depend on the resolution of the simulation, we carried out simulations at
a number of different resolutions: $60 \times 44$, $120 \times 88$, and $240 \times 176$ for initial
configuration A (see Table \ref{tab:tab1}), and $120 \times 88$ and $240 \times 176$ for initial
field configuration B (see Table \ref{tab:tab2}). The density and magnetic field structure in the
turbulent state is similar for all resolutions with initial configuration A. However, configuration B
requires a higher resolution to resolve the dynamics of the MRI. As a result, the lower resolution
simulation ($120 \times 88$) has a smaller $\langle \dot M \ra$ than the higher resolution simulation
($240 \times 176$).

As a dimensionless measure of whether our results for the RM have
converged, we define the coherence parameter (Coh) as \be
\label{eq:Coh} {\rm Coh} = \frac{\int_{r_{\rm in}}^{r_{\rm out}} n {\bf B \cdot dl} } {\int_{r_{\rm
in}}^{r_{\rm out}} n {\bf |B \cdot dl|}}. \ee Coh quantifies the
effect of cancellation due to magnetic field reversals on the inferred
$RM$. If the large-scale dynamics of the magnetic field is properly
resolved, one would expect that simulations with different resolution
would have statistically similar values for Coh.

Figure \ref{fig:fig14} shows the beam averaged Coh
at $\theta = 90^\circ$ (the equator) as a function of time for initial
configuration A for three different resolutions and $r_0 = 100 r_g$.
All three resolutions show statistically similar variations about
$\langle RM \ra \approx 0$ (corresponding to a similar average and
standard deviation for RM; see Table \ref{tab:tab1}). The time
variability is also quite similar, implying that the resolution does
not appear to be significantly affecting our determination of RM.

Figure \ref{fig:fig15} shows the beam averaged Coh
for initial configuration B as a function of time in the equatorial
plane for $r_0 = 100 r_g$.  The results are similar for the two
different resolutions ($240 \times 176$ and $120 \times 88$).  RM is
negative at most times because of the initial magnetic field.  There
are, however, times when RM is positive, consistent with the large
standard deviation shown in Figure \ref{fig:fig12}.

Ideally for Sgr A*, one would like to simulate an initial torus at
$r_0 \sim 10^5 r_g$, so that the inner disk extends to $\sim 10^3-10^4
r_g$, similar to the expected circularization radius in Sgr A* (e.g.,
\citealt{cua06}).  Because of finite computational resources, however,
we can only simulate a smaller disk.
We carried out simulations with $r_0=40 r_g~(r_{\rm out}=30 r_g),~100
r_g~(r_{\rm out}=60r_g)$, \& $200 r_g~(r_{\rm out}=120 r_g)$ for both
initial field configurations A \& B (see Tables \ref{tab:tab1} \&
\ref{tab:tab2}).  Although there are
some quantitative differences between the different simulations, the
basic results for RM, its statistical fluctuations, and its variation
with radius, do not seem to depend significantly on the radial extent
$r_0$, particularly when comparing the $r_0 = 100 r_g$ and $r_0 = 200
r_g$ simulations (the $r_0 = 40 r_g$ simulation has comparatively
little radial dynamic range). This provides some confidence in the
numerically calculated properties of RM through the accretion flow.

\section{Discussion}

We have carried out an analysis of Faraday rotation in global
axisymmetric MHD simulations of radiatively inefficient accretion
flows, motivated by the observations of linear polarization and
Faraday rotation in Sgr A*.  We first summarize our primary results
and then discuss their implications for models of Sgr A*.

We find that the accretion disk structure, and in particular the
rotation measure RM, retains memory of the initial magnetic field
threading the plasma, even in the quasi-steady turbulent state at late
times.  Physically, this is equivalent to a dependence on the
properties of the magnetic field in the plasma feeding the accretion
flow at large radii (which is typically hard to determine
observationally).  We considered two very different initial magnetic
field geometries, one with a net radial field in the equator of the
initial torus and one without (see Figs. \ref{fig:fig8}
[configuration B] \& \ref{fig:fig1} [configuration A], respectively).
In configuration B, the initial magnetic field unit vectors are
symmetric about the equatorial plane, while in configuration A they
are antisymmetric.  The initial symmetry of the magnetic field remains
in the RM as a function of viewing angle $\theta$ determined in the
turbulent accretion flow (compare Figs \ref{fig:fig5} \&
\ref{fig:fig12}).  In addition, the steady state density and $\beta$
profiles are rather different for the two different initial magnetic
field configurations (Figs. \ref{fig:fig10} \& \ref{fig:fig11}).
Given the anti-dynamo theorem, it is not that surprising that the
initial field structure imprints itself on the late-time disk
structure.  However, the similarity of previous 2D and 3D global disk
simulations \citep{haw01} suggests that this will also be true in
three dimensional simulations. It remains to be seen if
simulations of accretion disks in three dimensions lose memory of the
initial conditions at very late times (many orbital periods at large
radii), e.g., by undergoing an inverse cascade which generates a
large-scale magnetic field, or if the dependence on the large-scale
initial magnetic field that we see is a generic and robust feature of
accretion disk dynamos.

In spite of the dependence on the initial magnetic field threading the
plasma, we can draw a number of general conclusions about the
properties of Faraday rotation though magnetized geometrically thick
accretion flows.

\begin{enumerate}

\item{The RM through the midplane of the turbulent disk is highly time
dependent, with a typical dispersion in RM that is larger than the
mean. This implies significant changes in the observed sign of the RM.
We see sign changes in RM even for our initial condition with a net
radial field in the midplane of the torus (configuration B;
Fig. \ref{fig:fig12}), an initial condition that was chosen to
maximize the likelihood of a coherent sign of RM.}

\item{In the turbulent disk, RM receives roughly equal contributions
from all radii (Fig. \ref{fig:fig6} \& \ref{fig:fig13}).  Analytic
scaling arguments for disk models that are consistent with the
simulations also suggest a weak dependence of RM on radius, with small
radii dominating by a modest amount ($dRM/d\ln r \propto r^{-1/4}$; \S
\ref{sec:analytic}).} 

\item{RM is only a modest function of viewing angle $\theta$ (see Figs
\ref{fig:fig5} \& \ref{fig:fig12}); the polar RM is typically $\sim
0.2$ of the equatorial RM for $r_{\rm in}=10r_g$.}

\item{If there is a large scale magnetic field threading the plasma at
large radii (our configuration A), the resulting corona of the disk
also has a large-scale, stable magnetic field (see also SP01,
\citealt{dev06}).  The RM viewed through the corona has a dispersion
in time smaller than the mean (Fig. \ref{fig:fig5}) and is thus
relatively stable.}

\item{For polar viewing angles, the RM is primarily produced at small
radii; most of the contribution will thus arise from near the radius
at which the electrons become relativistic (Fig. \ref{fig:fig6} \&
\ref{fig:fig13}).}

\item{There is significant spatial variation in the amount of Faraday
rotation through a finite sized beam, with $\sigma_\perp(RM) \sim 0.3
\, RM$ at polar viewing angles (Fig. \ref{fig:fig7}).  This spatial
variation in RM can cause depolarization at frequencies for which
$\sigma_\perp(RM)\lambda^2 \gtrsim 1$.}

\end{enumerate}

\subsection{Implications for Sgr A*}

Our results confirm previous analytic arguments (e.g.,
\citealt{ago00,qua00}) that models of Sgr A* with $\dot M \sim \dot
M_{\rm Bondi} \sim 10^{-5} \mpy$ are ruled out by the measured Faraday
rotation of $RM \approx -6 \times 10^5 \, {\rm rad \, m^{-2}}$.  For
an accretion rate of $\dot M \sim \dot M_{\rm Bondi}$, the electrons
must be marginally relativistic even at small radii in order to
account for the low luminosity from Sgr A*.  In this case, our choice
of $r_{\rm in} = 10 r_g$ as the minimum radius for which Faraday
rotation is produced (see eq. [\ref{eq:RMcode}]) is quite reasonable
(and even conservative; in the ADAF models of \citealt{nar}, the
electron temperature never exceeds $6 \times 10^9$ K).  Our
simulations show that for nearly all viewing angles $\theta$ and for
both of the initial magnetic field configurations we considered, the
dimensionless rotation measure $RM_{\rm code} \gtrsim 10^{-4}$
(Figs. \ref{fig:fig5} \& \ref{fig:fig12}).  Even though the time
averaged $RM$ may be smaller than this for certain viewing angles
(e.g., through the equator of the turbulent disk, $\theta = 90^\circ$,
in Fig. \ref{fig:fig5}), the time variations are significant and so
$RM_{\rm code} \gtrsim 10^{-4}$ is a good indication of the typical
instantaneous RM that would be observed at any $\theta$.  Using
equation (\ref{eq:RM}) to convert RM to real units, we find that $\dot
M \sim \dot M_{\rm Bondi} \sim 10^{-5} \mpy$ would imply $RM \sim
10^{10} \rad$, over 4 orders of magnitude larger than the observed
value.  Thus such models are strongly ruled out.

We now discuss two scenarios that can account for the observed $RM
\approx - 6 \times 10^5 \, \rad$ and its apparent stability over the
past $\approx 7$ years.  These scenarios each identify the
RM-producing region with a key radial scale in the inflowing plasma.

\begin{figure}
\epsscale{2.2}
\centering \plottwo{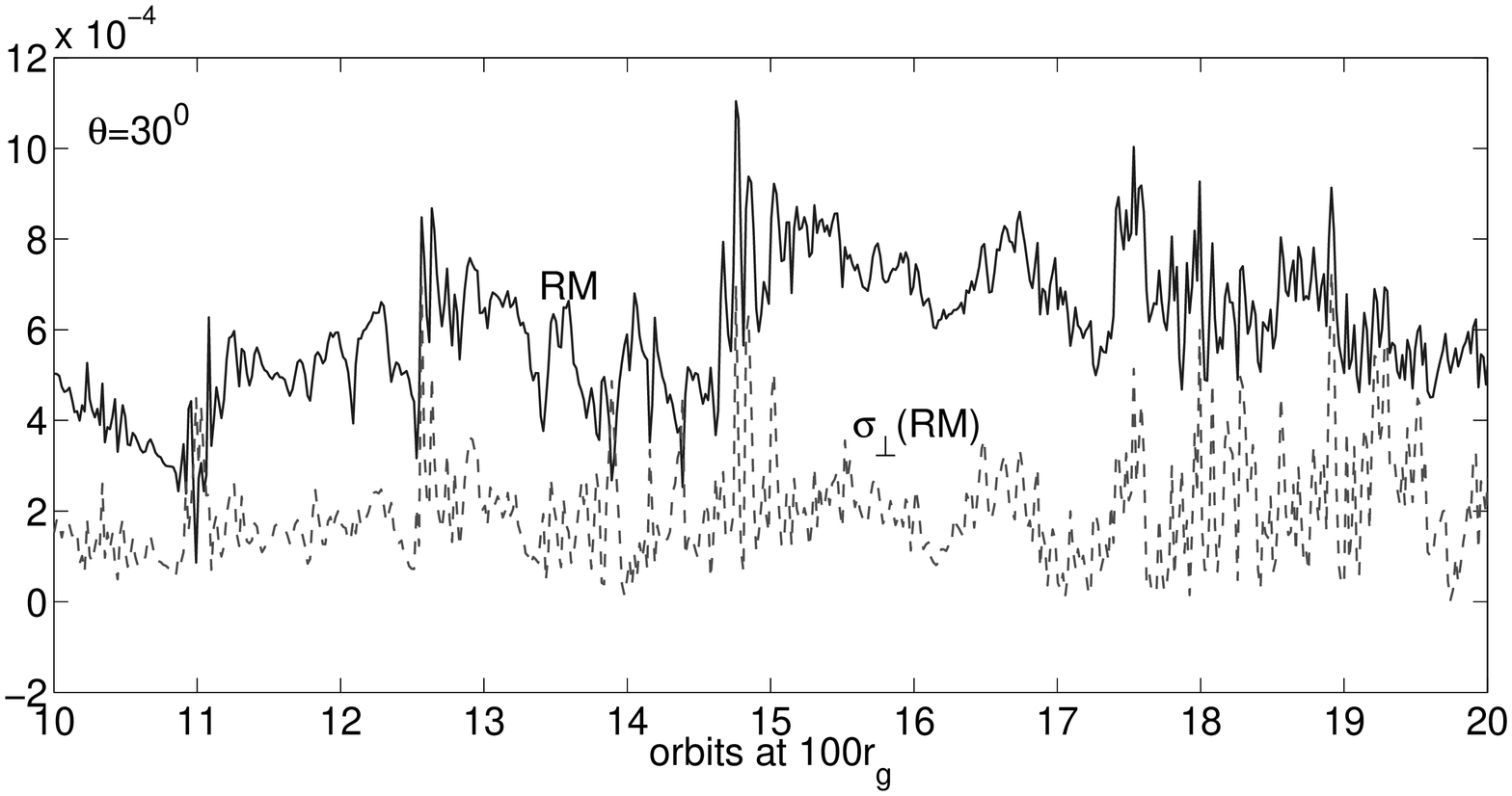}{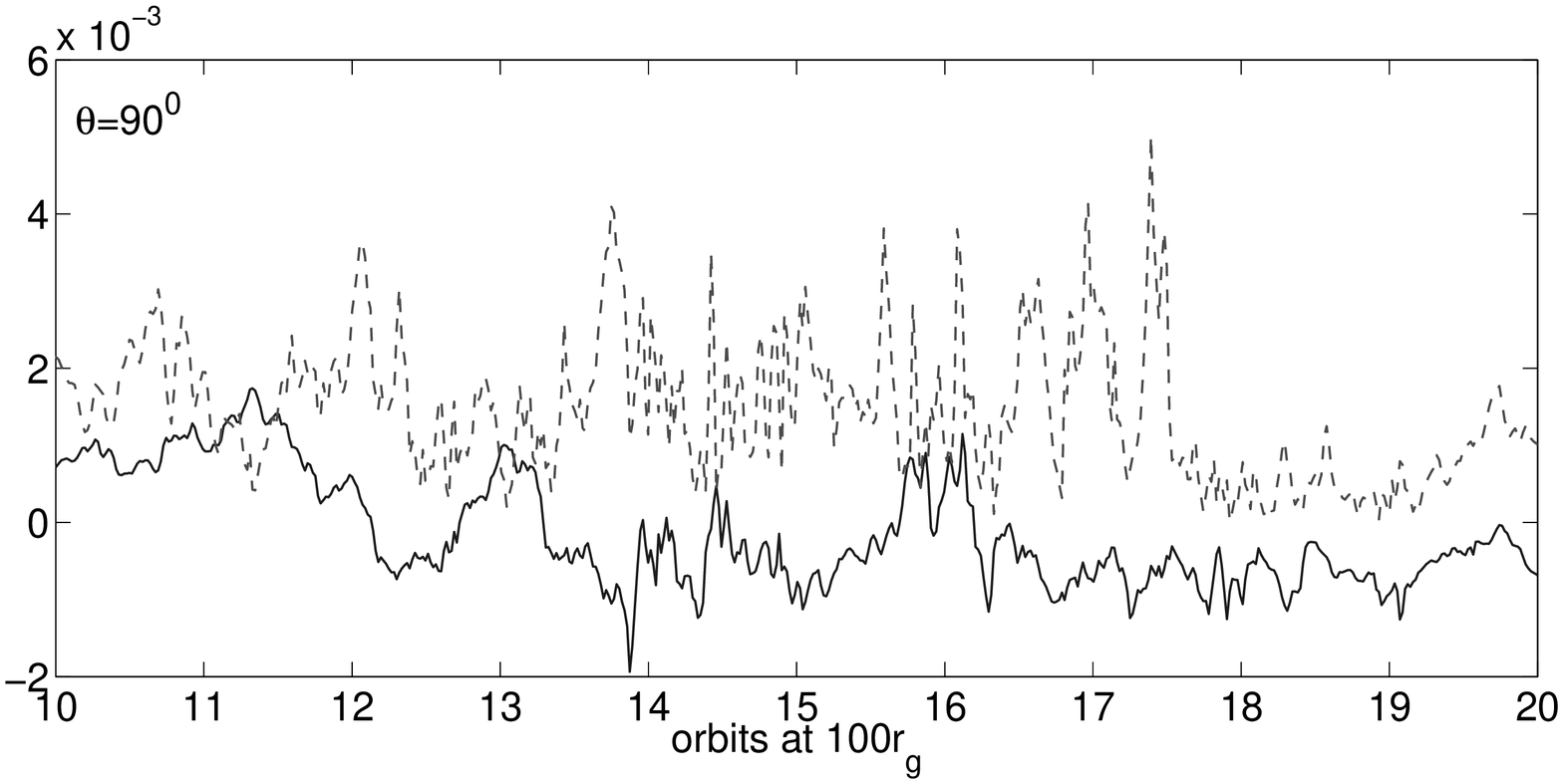}
\caption{Beam-averaged ($D_{\rm beam}=10r_g$) RM (solid line) and
standard deviation across the beam $\sigma_\perp(RM)$ (dashed line) as
a function of time for the fiducial simulation; one orbital period at
100 $r_g$ is $\approx 3 \, (M_{\rm BH}/4 \times 10^6 M_\odot)$ days.
Note that $\sigma_\perp(RM)$ in this Figure is a standard
deviation {\it accross the beam}, rather than a standard deviation
{\it in time}, which is shown in the remaining Figures.
{\it Top}: $\theta=30^\circ$; {\it Bottom}: $\theta=90^\circ$.  The
emission is beam depolarized when $\sigma_\perp(RM) \lambda^2 \gtrsim
1$.  Because the RM at polar viewing angles is primarily produced at
small radii (Fig. \ref{fig:fig6}), the characteristic timescale for RM
and $\sigma_\perp(RM)$ variability scales roughly as $r_{\rm
in}^{3/2}$; these calculations have $r_{\rm in} = 10 r_g$.
\label{fig:fig7}}
\end{figure}

First, our simulations show that if there is a relatively coherent
magnetic field in the plasma at large radii (configuration A) and if
the observer's line of sight is through the polar regions of the
accretion flow ($\theta \sim 30^\circ$), then the time variability of
RM appears to be sufficiently small to account for stable sign of RM
observed from Sgr A* (Figs. \ref{fig:fig5} \& \ref{fig:fig7}).  For
these polar viewing angles, we find $RM_{\rm code} \approx 3 \times
10^{-4}$, i.e., $RM \approx 4.5 \times 10^{10} \, (\dot M_{\rm in}/
10^{-5} \mpy)^{3/2} \, \rad$.  The RM at polar viewing angles is
produced at small radii with $d RM/d \ln r \propto r^{-\alpha}$ and
$\alpha \approx 1-1.5$.  Our simulations have $r_{\rm in} = 10 r_g$,
but $r_{\rm in} \sim 100 r_g$ is probably a more accurate estimate of
the radius at which the electrons become non-relativistic in Sgr A*
(e.g., \citealt{sha07}).  Rescaling our simulations to a larger value
of $r_{\rm in}$ yields $RM \approx 10^{9} \, (\dot M_{\rm in}/ 10^{-5}
\mpy)^{3/2} \, (r_{\rm in}/ 100 \, r_g)^{-1.5} \rad$ for $\alpha =
1.5$. Thus $RM \approx 6 \times 10^5 \, \rad$ requires $\dot M_{\rm
in} \approx 7 \times 10^{-8} \, (r_{\rm in}/ 100 r_g) \, \mpy$. For
$\alpha \approx 1$, which is also reasonably consistent with our
simulations, we find $\dot M_{\rm in} \approx 3 \times 10^{-8} \,
(r_{\rm in}/ 100 r_g)^{2/3} \, \mpy$. Because $r_{\rm in}$ is
only moderately well-constrained, this estimate of $\dot M_{\rm in}$
is probably uncertain by a factor of $\sim 5$.  In spite of this
uncertainty, our estimated $\dot M_{\rm in}$
required to account for the observed RM from Sgr A* is comparable to
previous analytic estimates (e.g., \citealt{mar07}); it is also
similar to the accretion rate estimate of \citet{sha07}, who inferred
$\dot M_{\rm in}$ by calculating the radiative efficiency of the
accreting plasma in Sgr A*.

A second possible explanation for the observed RM from Sgr A* is that
it is not directly produced within the nearly Keplerian accretion flow
at small radii, but is instead produced in the more slowly rotating
plasma further from the black hole.  Numerical simulations of
accretion onto Sgr A* from stellar winds located at $\sim 10^5-10^6
r_g$ find that the circularization radius of the flow is $r_{\rm circ}
\sim 10^3-10^4 r_g$ \citep{cua06}. As discussed in \S1, the RM
produced by the observed thermal plasma at $\sim 10^5 r_g$ is $\approx
5000 \, \beta^{-1/2} \, \rad$, where $\beta$ is the ratio of the
thermal energy density to the magnetic energy density, assuming a
magnetic field with a significant radial component.  If the Bondi
solution is applicable outside $r_{\rm circ}$, then $d RM/d \ln r
\propto r^{-7/4}$ and the RM produced by the roughly spherical
inflowing plasma at large radii is $RM \approx 10^7 \, \beta^{-3/2} \,
(r_{\rm circ}/ 10^3 r_g)^{-7/4} \, \rad$ (roughly independent of
viewing angle).  Circularization radii in the range expected can thus
plausibly account for the observed RM from Sgr A* if $\beta \sim 1$.
Note that in this interpretation, $\dot M_{\rm in}$ must be $\ll 7
\times 10^{-8} \, (r_{\rm in}/ 100 r_g) \, \mpy$ (for $\alpha = 1.5$)
in order for the Faraday rotation from $r \ll r_{\rm circ}$ to be
small compared to that from $\sim r_{\rm circ}$.

This interpretation requires a significant magnetic field at large
radii, with $\beta \sim 1$.  Such a field could perhaps be produced by
flux freezing of a moderately weaker field residing in the central
parsec $\sim 10^6 r_g$.  More interestingly, spherically inflowing
plasma in Sgr A* is unstable to the magnetothermal instability (MTI)
which amplifies magnetic fields in a low collisionality plasma when
the temperature increases in the direction of gravity
\citep{bal00,par05}.  The MTI preferentially operates at large radii
in spherical accretion flows ($\gtrsim 10^4 r_g$) because only there
is the growth time short compared to the inflow time (e.g., Fig. 9 of
\citealt{jq}).  In a hydrostatic atmosphere with a fixed temperature
difference between the top and bottom of the atmosphere, the MTI leads
to a significant radial heat flux, magnetic field amplification, and
field lines combed out in the radial direction \citep{par05}.
\citet{jq} argued that the heat flux associated with the MTI could
modify the dynamics of spherical accretion at large radii, which would
in turn alter $d RM/d \ln r$ from the Bondi estimate above.  Thus the
efficacy of the MTI for producing the observed Faraday rotation from
Sgr A* needs to be evaluated in a self-consistent dynamical
calculation.

The above interpretations of the Faraday rotation observed towards Sgr
A*, namely that it arises either from $r_{\rm in} \sim 100 r_g$ or $
r_{\rm circ} \sim 10^4 r_g$, can probably be distinguished by careful
monitoring of the variability of RM, which should be present at some
level even in the absence of sign changes.  In the $r_{\rm in}$
scenario one would expect variability on $\sim$ hour-day timescales,
while in the $r_{\rm circ}$ scenario, the variability would occur on
much longer timescales, of order years.  The apparent lack of
significant RM variability to date (e.g., \citealt{bow05,mar07}) may
favor models in which the RM is produced at large radii $\sim r_{\rm
circ}$.

If radii $\sim r_{\rm in}$ contribute significantly to the observed
RM, our simulations can also be used to quantify whether Faraday
depolarization can explain the rapid drop in the polarization fraction
observed at frequencies below a few 100
GHz~\citep[e.g.,][]{bow03,mar06}.  Beam depolarization becomes
important when $\sigma_\perp(RM) \lambda^2 \gtrsim 1$, where
$\sigma_\perp$ is the variation in RM across a finite-sized source
~\citep[e.g.,][]{bur66,tri91}.  Our simulations have $\sigma_\perp(RM)
\sim 0.1-1 \, RM$ for polar viewing angles, with the exact value
depending on $\theta$ and $r_{\rm in}$ (Fig. \ref{fig:fig7} and \S
\ref{sec:fid}).  Taking $\sigma_\perp(RM) \approx 0.3 \, RM$ as a
fiducial value, and using the observed RM of $\approx 6 \times 10^5
{\rm rad~m}^{-2}$, this implies that emission below $\approx 150$ GHz
should be depolarized by beam depolarization, in reasonable agreement
with observations. Because of the variations in $\sigma_\perp$
in our current simulations with $r_{\rm in}$ and $\theta$, higher
resolution simulations and simulations with $r_{\rm in} \sim 100 r_g$
will be required to more accurately determine the frequency below
which beam depolarization becomes important.
Figure \ref{fig:fig7} also shows that
$\sigma_\perp(RM)$ increases by a factor of $\sim 10$ at certain
times.  This could give rise to a significant decrease in the observed
polarization similar to that observed by \citet{mar06} at 340 GHz.

In a recent paper, \citet{bop} argue that the ``normal'' Faraday
rotation we have focused on in this paper may not be the dominant
process shaping the observed polarization from Sgr A*.  Instead, they
argue that because the plasma is relativistic over a large range of
radii, the normal modes of the plasma are elliptical.  In this case,
the observed polarization probes the properties of the normal modes in
the radio-emitting region, which in turn depends on the magnetic field
and temperature structure of the plasma.  However, current
observations of Sgr A* show that the change in the polarization angle
from 100-400 GHz is consistent with the $\lambda^2$ variation expected
for normal Faraday rotation (e.g., \citealt{mac06,mar07}). This
strongly suggests that non-relativistic plasma at large radii is
modifying the intrinsic polarization of the source via Faraday
rotation.  It also requires that the intrinsic polarization angle of
the source is relatively independent of frequency.  Finally, we note
that the source size is inferred to decrease significantly at the
frequencies for which linear polarization is detected, from $\approx
10 r_g$ at 100 GHz to $\approx 2 r_g$ at 400 GHz \citep{bow04}.  Thus
different frequencies sample different plasma along the line of sight
propagating to observers on Earth.  To produce the same RM independent
of frequency requires that the source size at all frequencies is much
smaller than the size of the region where the RM is produced (so that
there is no variation in RM with frequency).  This criterion is
reasonably satisfied for both of the interpretations of the Faraday
rotation from Sgr A* considered above.

Throughout this paper we have focused on the implications of our
results for Sgr A* because of the wealth of observational data
available.  We anticipate, however, that as observational techniques
improve our simulations will prove useful for interpreting the
physical conditions in the vicinity of other accreting black holes.

\acknowledgements

We thank Geoff Bower, Heino Falcke, and Dan Marrone for useful
conversations.  PS and EQ were supported in part by NASA grant
NNG06GI68G and the David and Lucile Packard Foundation. PS was partly
supported by DOE award DE-FC02-06ER41453 to Jon Arons. JS was
supported by DOE grant DE-FG52-06NA26217.  Most of the computing
resources were provided by the Princeton Plasma Physics Laboratory
Scientific Computing Cluster. This research also used resources of the
National Energy Research Scientific Computing Center, which is
supported by the Office of Science of the U.S. Department of Energy
under Contract No. DE-AC02-05CH11231.

\appendix
\section{Vector potential for configuration B}
\begin{figure}
\centering \plotone{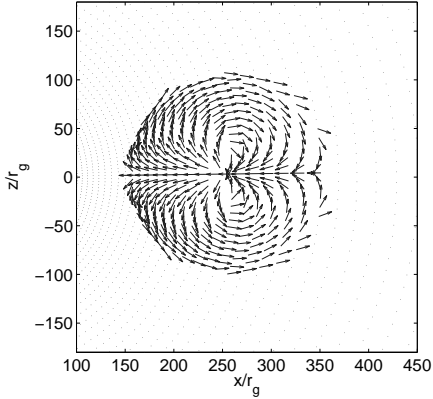} \caption{Initial magnetic field unit
vectors for field configuration B. In contrast to the initial magnetic
field for the fiducial simulations (Fig. \ref{fig:fig1}), in this case
the field is symmetric about the equatorial plane and there is a net
inward radial field in the equator. \label{fig:fig8}}
\end{figure}

\begin{figure}
\epsscale{1.11}
\centering \plottwo{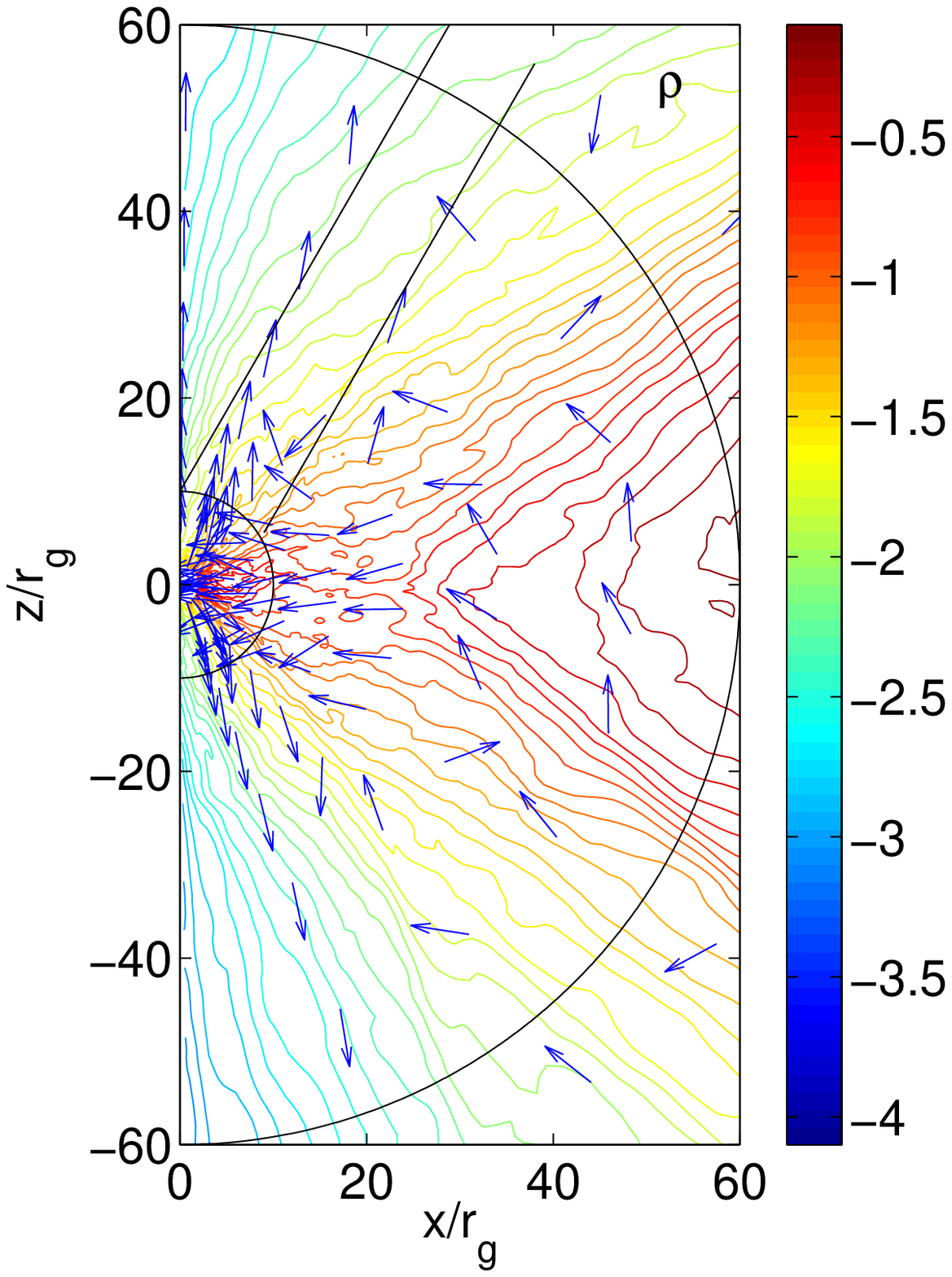}{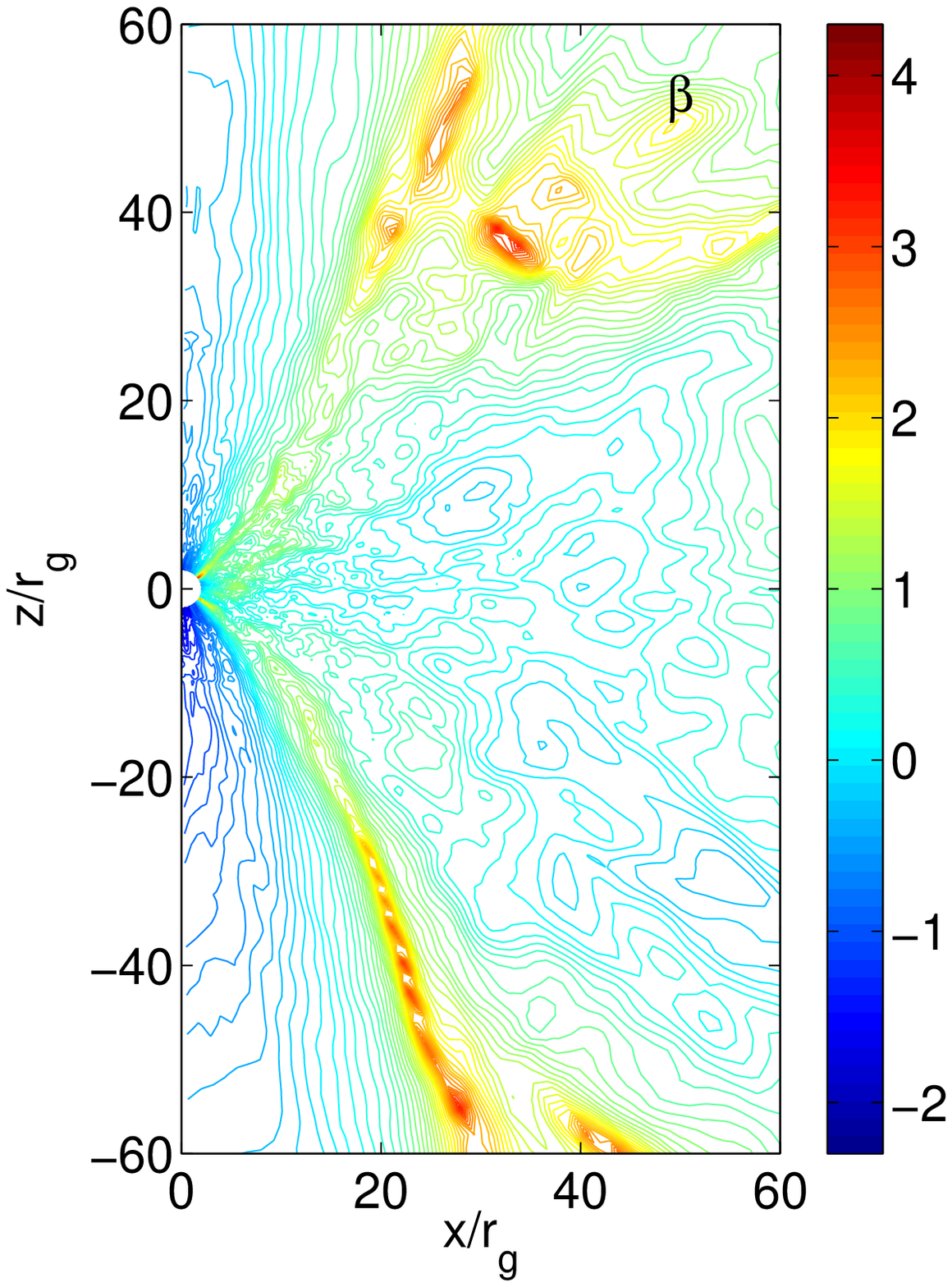}
\caption{Contour plots of density ($\rho$, left) and $\beta=8\pi
p/B^2$ (right) in the quasi-steady turbulent state for initial field
configuration B with $r_0 = 100 r_g$ and resolution $240 \times 176$
(averaged from 1.5 to 3.5 orbits at $100r_g$).  Arrows indicate the
average magnetic field direction.  The disk is less dense, more
strongly magnetized ($\beta \sim 1$), and much thicker than for the
fiducial simulation (see Fig. \ref{fig:fig3}).  The magnetic field in
the corona is less ordered than for the fiducial simulation, though
there is still very low $\beta$ plasma at the poles. The inner
($r_{\rm in}=10 r_g$) and outer ($r_{\rm out}=60 r_g$) radii used in
our rotation measure integrals are shown.  Also shown is a fiducial
``beam'' with a diameter of $D_{\rm beam}=10r_g$ at $\theta=30^\circ$
used to calculate the rotation measure for a finite sized
source. \label{fig:fig9}}
\end{figure}

\begin{figure}
\centering \plotone{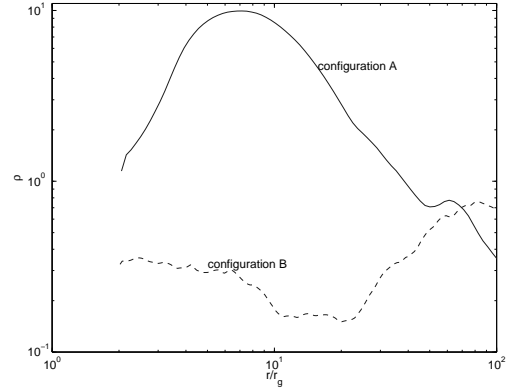} \caption{Density in the equatorial
plane as a function or radius, averaged over the turbulent state for
initial field configuration A (solid; Fig. \ref{fig:fig1}) and B
(dashed; Fig. \ref{fig:fig8}).  The density is averaged over a
$12^\circ$ wedge centered on the midplane. \label{fig:fig10}}
\end{figure}

\begin{figure}
\centering \plotone{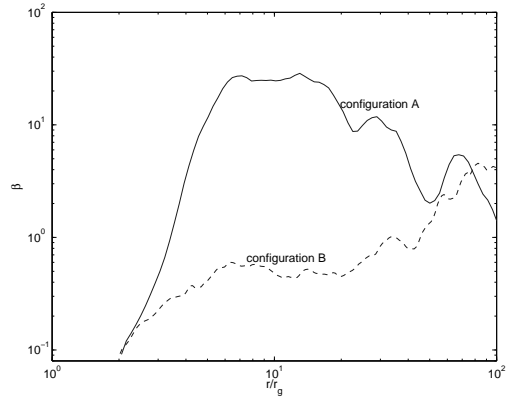} \caption{Plasma $\beta$ in the
equatorial plane as a function or radius, averaged over the turbulent
state for initial field configuration A (solid; Fig. \ref{fig:fig1})
and B (dashed; Fig. \ref{fig:fig8}); $\beta$ is averaged over a
$12^\circ$ wedge centered on the midplane.  The plasma $\beta$ is
significantly smaller for configuration B, likely because of flux
freezing of the initial radial field. \label{fig:fig11}}
\end{figure}

\begin{figure}
\centering \plotone{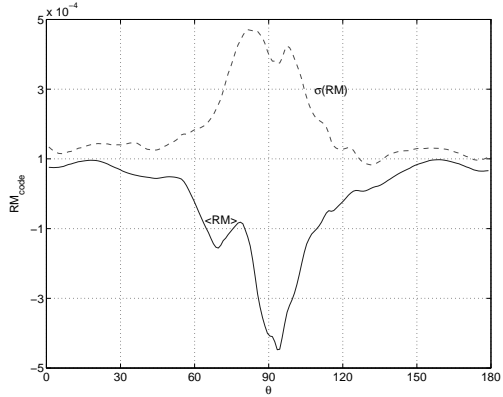} \caption{Time averaged RM
(solid line) and standard deviation in time (dashed line) in a $10
r_g$ diameter beam as a function of viewing angle for initial field
configuration B (for resolution $240 \times 176$, $r_{\rm in}=10r_g$,
$r_{\rm out}=60r_g$, and $r_0 = 100 r_g$).  The fluctuations in RM
exceed the mean at essentially all angles. \label{fig:fig12}}
\end{figure}

\begin{figure}
\centering \plotone{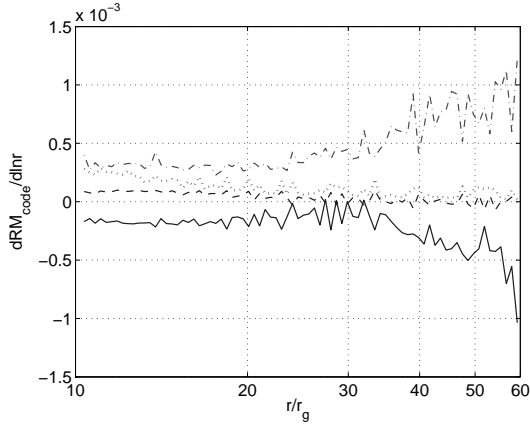} \caption{$dRM_{\rm code}/d\ln r$
for a $10 r_g$ beam for initial field configuration B (for resolution
$240 \times 176$, $r_{\rm in}=10r_g$, $r_{\rm out}=60r_g$, and $r_0 =
100 r_g$). Time averaged (solid line) and standard deviation in time
(dot-dashed line) for an equatorial viewing angle and time average
(dashed line) and standard deviation in time (dotted line) for a
$30^\circ$ viewing angle are shown. $dRM_{\rm code}/d\ln r$ decreases
with increasing radius for the $30^\circ$ viewing angle while it
increases with increasing radius for the equatorial viewing angle.
The curves are oscillatory in part because $dM_{\rm code}/d\ln r$
requires taking a numerical derivative. \label{fig:fig13}}
\end{figure}

\begin{figure}
\centering \plotone{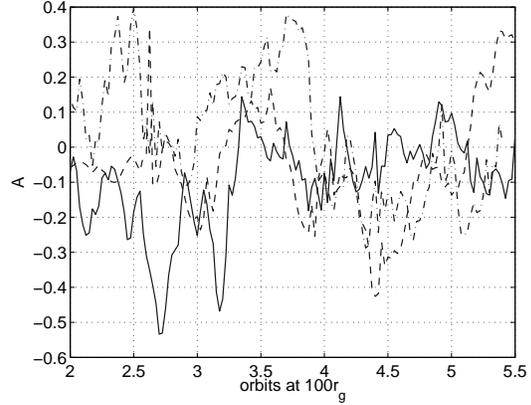}
\caption{The effect of magnetic field reversals on the rotation
measure is quantified by the Coherence (see eq.  [\ref{eq:Coh}]). Here
Coh is shown as a function of time for three different resolutions for
the fiducial run: $120 \times 88$ (solid line), $60 \times 44$ (dashed
line), and $240 \times 176$ (dot-dashed line).  An equatorial viewing
angle and a $10 r_g$ beam is assumed for these
calculations.\label{fig:fig14}}
\end{figure}

\begin{figure}
\centering \plotone{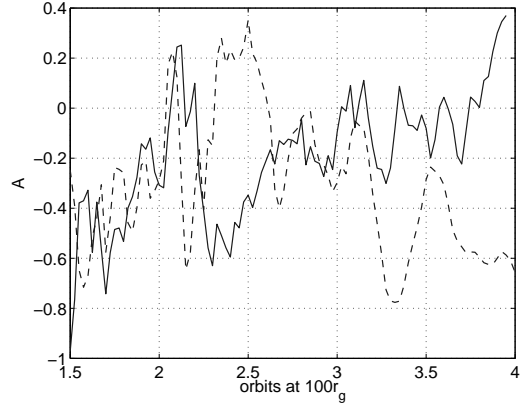}
\caption{Coherence (see eq. [\ref{eq:Coh}]) as a function of time for
initial field configuration B at two resolutions (with $r_{\rm
in}=10r_g$, $r_{\rm out}=60r_g$, and $r_0 = 100 r_g$): $120 \times
88$ (solid line) and $240 \times 176$ (dashed line). An equatorial
viewing angle and a $10 r_g$ beam is assumed for these calculations.
Note that the Coh is positive at some times, demonstrating that RM in
the midplane can change sign even though there is an initially radial
field threading the equator of the torus.
\label{fig:fig15}}
\end{figure}

To define the vector potential for initial field configuration B, we
set up a new cylindrical coordinate system whose origin is on the
equator of our standard spherical coordinate system at a distance of
$1.3 r_0$ (roughly the mid-point of the initial torus), and whose x 
and y axes are aligned with the local $-\theta$
and $-r$ axes of the spherical coordinate system, respectively. The
vector potential for initial field configuration B is then given by
\ba \nonumber A &=& 0.005 \cos \left ( \frac{\pi}{2} \frac{r_{\rm
cy}}{r_m} \right ) \cos(\phi) \hspace{1 in} {\rm for}~r_{\rm cy} <
r_m, \\ &=& 0 \hspace{2.42in} {\rm otherwise}, \ea where $r_{\rm cy}$
is the radius in the cylindrical coordinate system, $r_m=0.5 r_0$, and
$\phi$ is the angle with respect to the x- axis in the cylindrical
coordinate system. The vector potential is located at cell corners and
can be differenced to get the magnetic field unit vectors shown in
Figure \ref{fig:fig8}.

\clearpage

\begin{deluxetable}{ccccccccccc}
\tabletypesize{\scriptsize}
\rotate
\tablecaption{RM for Initial Field Configuration A \label{tab:tab1}}
\tablewidth{0pt}
\tablehead{
\colhead{$r_0$} & \colhead{Resolution} & \colhead{$\langle \dot{M}_{\rm code} \ra$} &
\colhead{$\langle{\rm Coh}\ra$\tablenotemark{a}} &
\colhead{$\sigma({\rm Coh})$}\tablenotemark{a} &
\colhead{$\langle RM_{code} \ra$}\tablenotemark{a} & \colhead{$\sigma(RM_{code})$}\tablenotemark{a} &
\colhead{$\langle{\rm Coh}\ra$}\tablenotemark{b} &
\colhead{$\sigma({\rm Coh})$}\tablenotemark{b} &
\colhead{$\langle RM_{code} \ra$}\tablenotemark{b} & \colhead{$\sigma(RM_{code})$}\tablenotemark{b} }
\startdata
200 $r_g$\tablenotemark{c} & 120 $\times$ 88 & 0.0025 & -0.1 & 0.22 & -1.1 $\times 10^{-4}$  & 4.1 $\times 10^{-4}$ & 0.85 & 0.13 & 2.8 $\times 10^{-4}$  & $ 10^{-4}$ \\
200 $r_g$ & 60 $\times$ 44 & 0.002 &0.04 & 0.16 & 8.3 $\times 10^{-5}$ & 2.7 $\times 10^{-4}$ & 0.95 & 0.088 & 2.1 $\times 10^{-4}$ & $10^{-4}$ \\
100 $r_g$ & 120 $\times$ 88 & 0.0029 & -0.05 & 0.14 & -1.6 $\times 10^{-4}$ & 3.9 $\times 10^{-4}$ & 0.83 & 0.14 & 1.6 $\times 10^{-4}$ & 5.7 $\times 10^{-5}$ \\
100 $r_g$ & 60 $\times$ 44 & 0.0022 & -0.04 & 0.17 & -8.5 $\times 10^{-5}$ & 2.6 $\times 10^{-4}$ & 0.93 & 0.12 & 2.3 $\times 10^{-4}$ & 6.8 $\times 10^{-5}$ \\
100 $r_g$ & 240 $\times$ 176 & 0.0033 & 0.04 & 0.18 & 8 $\times 10^{-5}$ & 3.4 $\times 10^{-4}$ & 0.63 & 0.26 & 1.4 $\times 10^{-4}$ & 9 $\times 10^{-5}$ \\
40 $r_g$ & 120 $\times$ 88 & 0.0031 & -0.06 & 0.12 & -2.2 $\times 10^{-4}$ & 3.7 $\times 10^{-4}$ & 0.66 & 0.24 & 9.5 $\times 10^{-5}$ & 5.7 $\times 10^{-5}$ \\
40 $r_g$ & 60 $\times$ 44 & 0.0028 & -0.21 & 0.17 & -4.2 $\times 10^{-4}$ & 3.1 $\times 10^{-4}$ & 0.84 & 0.15 & 1.9 $\times 10^{-4}$ & 8.1 $\times 10^{-5}$ \\
40 $r_g$ & 240 $\times$ 176 & 0.0039 & -0.03 & 0.1 & -1.4 $\times 10^{-4}$ & 4 $\times 10^{-4}$ & 0.69 & 0.22 & 8.5 $\times 10^{-5}$ & 5.8 $\times 10^{-5}$ \\
\enddata
\tablecomments{$\langle \ra$ ($\sigma$) represents a time average (temporal standard deviation) in the turbulent
state. The time average is taken over the quasi-steady turbulent state which lasts from $\approx 3-10$ orbits at $r_0$ depending on the simulation.  All RM results are also averaged over a $10 r_g$ beam.}
\tablenotetext{a}{$\theta=90^\circ$}
\tablenotetext{b}{$\theta=30^\circ$}
\tablenotetext{c}{The fiducial run.}
\end{deluxetable}

\begin{deluxetable}{ccccccccccc}
\tabletypesize{\scriptsize}
\rotate
\tablecaption{RM for Initial Field Configuration B\label{tab:tab2}}
\tablewidth{0pt}
\tablehead{
\colhead{$r_0$} & \colhead{Resolution} & \colhead{$\langle \dot{M}_{\rm code} \ra$} &
\colhead{$\langle{\rm Coh}\ra$}\tablenotemark{a} &
\colhead{$\sigma({\rm Coh})$}\tablenotemark{a} &
\colhead{$\langle RM_{code} \ra$}\tablenotemark{a} & \colhead{$\sigma(RM_{code})$}\tablenotemark{a} &
\colhead{$\langle{\rm Coh}\ra$}\tablenotemark{b} &
\colhead{$\sigma({\rm Coh})$}\tablenotemark{b} &
\colhead{$\langle RM_{code} \ra$}\tablenotemark{b} & \colhead{$\sigma(RM_{code})$}\tablenotemark{b} }
\startdata
200 $r_g$ & 120 $\times$ 88 & 0.0016 & -0.17 & 0.27 & -5.9 $\times 10^{-4}$ & 9.1 $\times 10^{-4}$ & 0.45 & 0.29 & 7.7 $\times 10^{-5}$ & 9 $\times 10^{-5}$ \\
100 $r_g$ & 120 $\times$ 88 & 0.0031 & -0.1 & 0.25 & -3.4 $\times 10^{-4}$ & 8.8 $\times 10^{-4}$ & 0.64 & 0.25 & 6.9 $\times 10^{-5}$ & 4.6 $\times 10^{-5}$ \\
100 $r_g$ & 240 $\times$ 176 & 0.0061 & -0.2 & 0.21 & -3.4 $\times 10^{-4}$ & 3.8 $\times 10^{-4}$ & 0.15 & 0.45 & 3.2 $\times 10^{-5}$ & $ 10^{-4}$ \\
40 $r_g$ & 120 $\times$ 88 & 0.0059 & -0.29 & 0.17 & -0.0012 & 7.2 $\times 10^{-4}$ & 0.47 & 0.33 & 1.2 $\times 10^{-5}$ & 1.9 $\times 10^{-5}$ \\
40 $r_g$ & 240 $\times$ 176 & 0.0095 & -0.01 & 0.12 & -7.1 $\times 10^{-6}$ & 4.6 $\times 10^{-4}$ & 0.13 & 0.35 & $ 10^{-5}$ & 2.6 $\times 10^{-5}$ \\
\enddata
\tablecomments{$\langle \ra$ ($\sigma$) represents a time average (temporal standard deviation) in the turbulent
state. The time average is taken over the quasi-steady turbulent state which lasts from $\approx 3-10$ orbits at $r_0$ depending on the simulation. All RM results are also averaged over a $10 r_g$ beam.}
\tablenotetext{a}{$\theta=90^\circ$}
\tablenotetext{b}{$\theta=30^\circ$}
\end{deluxetable}

\clearpage

\epsscale{1}

\epsscale{1}

\epsscale{1}

\epsscale{1.}

\end{document}